\documentclass[showpacs,aps,graphicx]{revtex4}

\usepackage{graphicx}
\usepackage{amsmath}
\usepackage{amssymb}
\begin{document}

\title{High-efficiency multipartite entanglement purification of electron-spin states with charge
detection\footnote{Published in Quant. Inform. Proc.  \textbf{12},
855 - 876 (2013). (DOI 10.1007/s11128-012-0427-2)}}

\author{Tao Li, Bao-Cang Ren, Hai-Rui Wei, Ming Hua, and Fu-Guo Deng\footnote{Corresponding author. E-mail address: fgdeng@bnu.edu.cn.}}
\address{Department of Physics, Applied Optics Beijing Area Major Laboratory,  Beijing Normal University, Beijing 100875, China}

\date{\today }

\begin{abstract}
We present a high-efficiency  multipartite entanglement purification
protocol (MEPP) for electron-spin systems in a
Greenberger-Horne-Zeilinger state based on their spins and their
charges. Our MEPP contains two parts. The first part is our normal
MEPP with which the parties can obtain a high-fidelity $N$-electron
ensemble directly, similar to the MEPP with controlled-not gates.
The second one is our recycling MEPP with entanglement link from
$N'$-electron subsystems  ($2 < N' < N$). It is interesting to show
that the $N'$-electron subsystems can be obtained efficiently by
measuring the electrons with potential bit-flip errors from the
instances which are useless and are just discarded in all existing
conventional MEPPs. Combining these two parts, our MEPP has the
advantage of the efficiency higher than other MEPPs largely for
electron-spin systems.
\end{abstract}


\maketitle

\section{introduction}

Entanglement has been regarded as  the  essential resource for
quantum information processing  and quantum communication
\cite{book}. From a perspective of applications, the entangled
systems shared by the separated parties in distant locations  are
required to be in a maximally entangled state for the efficiency and
the security of quantum communication
\cite{Ekert91,BBM92,LongLiu,teleportation,densecoding,densecoding2}.
Especially, multipartite entangled systems  have  many vital
applications in quantum  computation \cite{book} and quantum
communication, such as controlled teleportation
\cite{cteleportation,cteleportation2},  quantum state sharing
\cite{QSTS,QSTS1,QSTS2}, quantum secret sharing
\cite{QSS,QSS2,QSS3}, and so on. However, all these tasks are based
on the fact that the quantum channel with multipartite entangled
states shared by the legitimate distant participants has been set up
beforehand.   It is well known that the parties in quantum
communication cannot create nonlocal entanglement with local
operations and classical communication (LOCC). The distribution of
entanglement created locally  is inevitable. However, in a practical
transmission, the particles propagated away from each other are
destined to  suffer from channel noises, which will degrade the
entanglement or even make the maximally entangled state become a
mixed one. Therefore, it will decrease the fidelity of quantum
teleportation \cite{teleportation} and  quantum dense coding
\cite{densecoding,densecoding2}, and make the quantum communication
insecure \cite{Ekert91,BBM92,LongLiu}.

Recently, much attention has been drawn to entanglement purification
\cite{Bennett1,Deutsch,Pan1,Simon,shengpra,Murao,Horodecki,Yong,shengepjd,dengEMEPP,shengpratwostep,lixhepp,shengpraonestep,dengonestep,wangcqic},
a fascinating tool for the parties in  quantum communication to
extract some high-fidelity entangled states from a set of less
entangled systems. The original entanglement purification protocol
(EPP) by Bennett \emph{et al.} \cite{Bennett1} and that by  Deutch
\emph{et al.} \cite{Deutsch} are expressed in terms of  the quantum
controlled-not (CNOT) logic operations. Subsequently, Pan \emph{et
al.} \cite{Pan1} introduced an EPP with linear optical elements
based on the polarization degree of freedom of photons.  In 2002,
Simon and Pan \cite{Simon} presented an EPP with a currently
available parametric down-conversion (PDC) source. In 2008, Sheng
\emph{et al.} \cite{shengpra} proposed an efficient EPP based on a
PDC source with cross-Kerr nonlinearity and it can, in principle, be
repeated to obtain a high-fidelity entangled ensemble. In 2011,
Wang, Zhang, and Jin \cite{wangcqic} proposed an interesting EPP
based on cross-Kerr nonlinearity and the measurement on the
intensity of coherent beams. In 2010, Sheng and Deng
\cite{shengpratwostep} introduced the concept of deterministic
entanglement purification and proposed a two-step deterministic
entanglement purification protocol (DEPP), the first DEPP in which
the parties can obtain a maximally entangled state from each system
transmitted, far different from the conventional entanglement
purification protocols (CEPPs)
\cite{Bennett1,Deutsch,Pan1,Simon,shengpra,wangcqic,Murao,Horodecki,Yong,shengepjd,dengEMEPP}.
Subsequently, a one-step DEPP
\cite{shengpraonestep,lixhepp,dengonestep} was proposed, only
resorting to the spatial entanglement or the frequency entanglement
of a practical PDC source and linear optical elements. In essence,
both the CEPPs
\cite{Bennett1,Deutsch,Pan1,Simon,shengpra,wangcqic,Murao,Horodecki,Yong,shengepjd,dengEMEPP}
and the DEPPs
\cite{shengpratwostep,shengpraonestep,lixhepp,dengonestep} are based
on  entanglement transfer. Taking the EPP for photon systems as an
example, the CEPPs are based on the entanglement transfer between
different entangled photon systems, while the DEPPs are based on the
transfer between different degrees of freedom of the entangled
photon system itself. The DEPPs require that at least one degree of
freedom of photons is stable when the photons are transmitted over a
noisy channel, while the CEPPs only require that there is
entanglement in the ensemble after transmission.

Up to now,  most of EPPs
\cite{Bennett1,Deutsch,Pan1,Simon,shengpra,shengpratwostep,lixhepp}
are focused on bipartite entangled photon systems and there are only
several multipartite entanglement purification protocols, including
high-dimension EPPs \cite{Murao,Horodecki,Yong,shengepjd,dengEMEPP}.
For instance,  in 1998, Murao \emph{et al.} \cite{Murao} proposed
the first multipartite entanglement purification protocol (MEPP) to
purify multipartite entangled systems in a Werner-type state, with
CNOT gates. In 2007, this protocol was extended to high-dimensional
multipartite quantum systems by Cheong \emph{et al.} \cite{Yong},
resorting to some generalized XOR gates, instead of the common CNOT
gates. In 2009, with the considerable experimental progress
achieved, Sheng \emph{et al.} introduced a feasible  MEPP for
$N$-photon systems in a Greenberger-Horne-Zeilinger (GHZ) state
\cite{shengepjd}. In their protocol, quantum nondemolition detector
(QND) is exploited to fulfill the functions of the parity-check
gate. With QNDs, the parties can obtain some high-fidelity GHZ-state
systems from the less entangled ones by performing the protocol
iteratively.

A conduction electron can act as a qubit in both the charge degree
of freedom and the spin degree of freedom, which are relatively
independent on each other. In other words, when we measure the
charge of an electron system, its spin state  would be kept
unaffected, and vice versa. Owing to this fantastic feature, charge
detection \cite{cd} has been exploited to accomplish many works,
such as the CONT gate between electronic qubits \cite{beenakker},
the generation of the entangled spins \cite{paritybox}, the
multipartite entanglement analyzer \cite{cluster}, and so on. Also,
some EPPs and an entanglement concentration protocol for electron
systems have been proposed \cite{feng,shengpla,wangcpra,shengplaec}.
For example, in 2005, Feng \emph{et al.} \cite{feng} proposed an
electronic EPP for purifying two-electron systems in  a Werner state
with parity-check measurements based on charge detection
\cite{beenakker}, following some ideas in the original EPP proposed
by Bennett \emph{et al.} \cite{Bennett1} for photon pairs. In 2011,
Sheng \emph{et al.} \cite{shengpla} presented a MEPP for
electron-spin states and Wang \emph{et al.} \cite{wangcpra} proposed
a two-electron EPP by using quantum-dot spins in optical
microcavities. Although  there are some MEPPs and some two-electron
EPPs,  the efficiency in these protocols is relatively low.

In this work, we present a high-efficiency MEPP for electron-spin
systems in a GHZ state with charge detection. It contains two parts.
One is our normal MEPP with which the parties in quantum
communication can distill a high-fidelity $N$-electron ensemble
directly, by replacing perfect CNOT gates with the parity- check
detectors based on charge detection in Ref. \cite{Murao}, but with a
higher efficiency. The other is our recycling MEPP in which the
entanglement link based on charge detection is used to produce some
$N$-electron entangled systems from entangled $N'$-electron
subsystems ($2 \leq N' < N$). It is interesting to show that the
entangled $N'$-electron subsystems can be obtained  efficiently from
the cross-combination items, which are useless and are just
discarded in all existing conventional EPPs
\cite{Bennett1,Deutsch,Pan1,Simon,shengpra,wangcpra,wangcqic,feng,Murao,Horodecki,Yong,shengepjd,shengpla}.
With these two parts, the present MEPP  for electron-spin states has
efficiency higher than all other  MEPPs largely. We discuss the
detail of our high-yield MEPP for electron-spin states of
three-electron systems and its principle is suitable to the
$N$-electron systems in an arbitrary GHZ state.

\section{High-efficiency three-electron entanglement purification for bit-flip errors with entanglement link and charge detection}    

Our thee-electron EPP contains two parts, which makes it different
from others
\cite{Bennett1,Deutsch,Pan1,Simon,shengpra,wangcpra,wangcqic,feng,Murao,Horodecki,Yong,shengepjd,shengpla}.
One is our normal MEPP with which the parties can obtain a
high-fidelity three-electron ensemble directly, similar to all
existing MEPPs. The other is our recycling MEPP with entanglement
link from subsystems. In the second part, the quantum resources are
obtained from the systems with less entanglement which are just
discarded   in all other MEPPs. We introduce the principles of these
two parts independently as follows.

\subsection{ Normal three-electron entanglement purification for bit-flip errors}

\begin{figure}[!h]
\begin{center}
\includegraphics[width=7cm,angle=0]{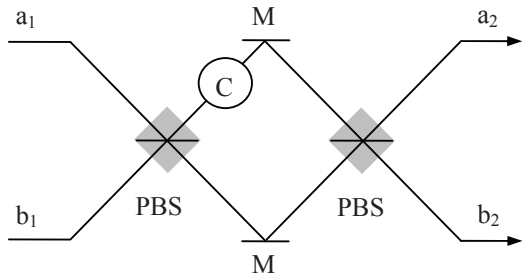}
\caption{The principle of a parity-check detector (PCD) based on
charge detection. PBS represents a polarizing beam splitter for
electron spins, which transmits the electron in the spin-up state
$|\uparrow\rangle$ and reflects the electron in the spin-down state
$|\downarrow\rangle$, respectively.  $M$ represents a mirror for the
spins of an electron. $C$ represents a charge detector and it can
distinguish the electron number $C=1$ from $C=0$.}\label{fig1}
\end{center}
\end{figure}

Before we start to explain the principle of our MEPP, we  present a
detailed description of a parity-check detector (PCD) which is
thought to be more feasible as a basic element for EPP than a
perfect CNOT gate. In Fig.\ref{fig1}, the polarizing beam splitter
(PBS) transmits the electrons in the spin-up state
$|\uparrow\rangle$ and reflects the ones in the spin-down state
$|\downarrow\rangle$. Therefore, for two electrons coming from two
different inputs of the first PBS, if they leave through different
outputs of the first PBS, the  charge detector (C) will get the
charge occupation number $C=1$; otherwise $C=0$ or 2. The charge
detector can distinguish the occupation number one from the
occupation numbers 0 and 2, but it cannot distinguish the case
between 0 and 2. In other words, it can only distinguish the
instances that the occupation number is even or odd
\cite{beenakker}. With this feature, one can see that the states
$|\uparrow\uparrow\rangle$ and $|\downarrow\downarrow\rangle$ will
lead the charge detector to obtain the charge occupation number
$C=1$ as the two  electrons passing through the first PBS will leave
through different modes. However, the states
$|\uparrow\downarrow\rangle$ and $|\downarrow\uparrow\rangle$ will
lead the charge detector to be $C=0$ and $C=2$, respectively. As
mentioned above, the charge detector cannot distinguish 0 and 2, and
it will show the same result, i.e., $C=0$ for simplicity. That is,
we can distinguish the states $|\uparrow\uparrow\rangle$ and
$|\downarrow\downarrow\rangle$ from $|\uparrow\downarrow\rangle$ and
$|\downarrow\uparrow\rangle$, according to  the different outcomes
of the charge detector.  The second PBS is used to split the two
electrons passing through the charge detector, without destroying
their spin states. In essence, the setup shown in Fig.\ref{fig1} is
just a parity-check detector (PCD) for the spins of two electrons.

For a three-electron spin system, there are eight GHZ states,
\begin{eqnarray}
|\Phi_{0}^{\pm}\rangle_{ABC}=\frac{1}{\sqrt{2}}(|\uparrow\uparrow\uparrow\rangle
\pm|\downarrow\downarrow\downarrow\rangle)_{ABC},\nonumber\\
|\Phi_{1}^{\pm}\rangle_{ABC}=\frac{1}{\sqrt{2}}(|\downarrow\uparrow\uparrow\rangle
\pm|\uparrow\downarrow\downarrow\rangle)_{ABC},\nonumber\\
|\Phi_{2}^{\pm}\rangle_{ABC}=\frac{1}{\sqrt{2}}(|\uparrow\downarrow\uparrow\rangle
\pm|\downarrow\uparrow\downarrow\rangle)_{ABC},\nonumber\\
|\Phi_{3}^{\pm}\rangle_{ABC}=\frac{1}{\sqrt{2}}(|\uparrow\uparrow\downarrow\rangle
\pm|\downarrow\downarrow\uparrow\rangle)_{ABC}.                                   \label{GHZstate}
\end{eqnarray}
Here the subscripts $A$, $B$, and $C$ represent the three electrons
belonging to the three parties, say Alice, Bob, and Charlie,
respectively. Suppose that the original GHZ state transmitted is
$|\Phi_{0}^{+}\rangle_{ABC}$. As we know, the noisy channel will
inevitably  degrade the entanglement of the state or even make it
be a mixed one. In detail, if the initial state
$|\Phi_{0}^{+}\rangle_{ABC}$ becomes $|\Phi_{i}^{+}\rangle_{ABC}$,
 a bit-flip error takes place on the $i$-th qubit ($i=1,2,3 $). If the state  $|\Phi_{0}^{+}\rangle_{ABC}$ evolves to $|\Phi_{0}^{-}\rangle_{ABC}$,
we say that  a phase-flip error appears. Sometimes, both a bit-flip
error and a phase-flip error will take place on the three-electron
system  such as the state $|\Phi_{i}^{-}\rangle_{ABC}$.  In order to
purify three-electron entangled systems, we are required to correct
both bit-flip errors and phase-flip errors on the quantum system.
Usually, an EPP can be divided into two steps
\cite{Pan1,shengpla,shengpra}. One is used to purify the bit-flip
error and the other is to the phase-flip error. In the second step,
the phase-flip error will be transformed into  the bit-flip error
with a Hadamard operation on each qubit and then the parties purify
the bit-flip error with the similar processes to these in the first
step. That is, the phase-flip error can, in principle,  be purified
with the similar processes \cite{Pan1}. We only discuss the
principle of the present MEPP for three-electron systems with
bit-flip errors below.

\begin{figure}[!h]
\begin{center}
\includegraphics[width=8cm,angle=0]{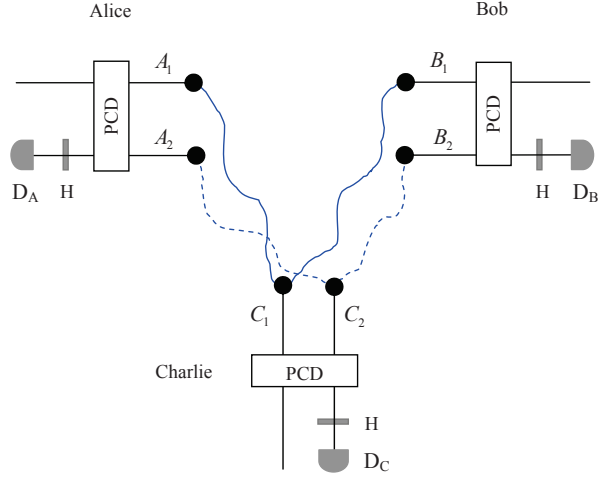}
\caption{The principle of our normal three-electron EPP with PCDs
based on charge detection. PCD represents a parity-check detector.
$H$ represents a Hadamard operation. $D_A$, $D_B$, and $D_C$
represent the single-electron measurements with the basis $Z=\{\vert
\uparrow\rangle, \vert \downarrow\rangle\}$ done by Alice, Bob, and
Charlie, respectively.} \label{fig2_MEEPP}
\end{center}
\end{figure}

Suppose the state of the tripartite electronic systems $\rho$ shared
by Alice, Bob, and Charlie is
\begin{eqnarray}
\rho_{ABC} &=&
F_{0}|\Phi_{0}^{+}\rangle\langle\Phi_{0}^{+}|+F_{1}|\Phi_{1}^{+}\rangle\langle\Phi_{1}^{+}|
+ F_{2}|\Phi_{2}^{+}\rangle\langle\Phi_{2}^{+}| +
F_{3}|\Phi_{3}^{+}\rangle\langle\Phi_{3}^{+}|.\label{ensemblerho}
\end{eqnarray}
Here $F_{0}$ is the fidelity of the state $|\Phi_{0}^{+}\rangle$
after it is transmitted over a noisy channel.  $F_{i}$ ($i=1,2,3$)
is the probability that the three-electron system is in the state
$|\Phi_{i}^{+}\rangle$. They satisfy the relation
\begin{eqnarray}
F_{0}+F_{1}+F_{2}+F_{3}=1.
\end{eqnarray}
For obtaining some high-fidelity three-electron entangled systems,
the three parties should operate a pair of three-electron systems in
the state  $\rho$ with  LOCC. The principle of our normal
three-electron EPP is shown in Fig.\ref{fig2_MEEPP}. We label the
two three-electron systems with $A_{1}B_{1}C_{1}$ and
$A_{2}B_{2}C_{2}$, respectively. The state of the six-electron
system $A_{1}B_{1}C_{1}A_{2}B_{2}C_{2}$ is $\rho_{A_1B_1C_1}\otimes
\rho_{A_2B_2C_2}$. It can be viewed as the mixture of the 16 pure
states, i.e., $|\Phi_{i}^{+}\rangle\otimes|\Phi_{j}^{+}\rangle$ with
the probability of $F_{i}F_{j}$ ($i,j=0,1,2,3$). The three parties
make the electron pair they own pass through their PCDs. That is,
the electron $A_{1}$ entrances the up spatial mode and $A_{2}$ the
down-spatial mode, for comparing the spin parity of their electron
pair. After the parity-check measurements, Alice, Bob, and Charlie
communicate their outcomes, and they keep the instances in which all
the three parties obtain an even parity or an odd parity.

When the parities of  the electron pairs obtained by Alice, Bob, and
Charlie are all even, the state of the complicated system composed
of the six electrons $A_{1}B_{1}C_{1}A_{2}B_{2}C_{2}$ becomes a
mixed one $\rho_{t_{even}}$ (without normalization),
\begin{eqnarray}
\rho_{t_{even}} &=&
\frac{1}{2}(F_{0}^{2}|\phi_{0}\rangle\langle\phi_{0}|+F_{1}^{2}|\phi_{1}\rangle\langle\phi_{1}|+
F_{2}^{2}|\phi_{2}\rangle\langle\phi_{2}|
+F_{3}^{2}|\phi_{3}\rangle\langle\phi_{3}|).\label{ensemblerho1}
\end{eqnarray}
where
\begin{eqnarray}
|\phi_{0}\rangle &=&
\frac{1}{\sqrt{2}}(|\uparrow\uparrow\uparrow\rangle_{A_1B_1C_1}|\uparrow\uparrow\uparrow\rangle_{A_{2}B_{2}C_{2}}
+
|\downarrow\downarrow\downarrow\rangle_{A_{1}B_{1}C_{1}}|\downarrow\downarrow\downarrow\rangle_{A_{2}B_{2}C_{2}}),
\end{eqnarray}
\begin{eqnarray}
|\phi_{1}\rangle &=&
\frac{1}{\sqrt{2}}(|\downarrow\uparrow\uparrow\rangle_{A_{1}B_{1}C_{1}}|\downarrow\uparrow\uparrow\rangle_{A_{2}B_{2}C_{2}}
+
|\uparrow\downarrow\downarrow\rangle_{A_{1}B_{1}C_{1}}|\uparrow\downarrow\downarrow\rangle_{A_{2}B_{2}C_{2}},
\end{eqnarray}
\begin{eqnarray}
|\phi_{2}\rangle &=&
\frac{1}{\sqrt{2}}(|\uparrow\downarrow\uparrow\rangle_{A_{1}B_{1}C_{1}}|\uparrow\downarrow\uparrow\rangle_{A_{2}B_{2}C_{2}}
+
|\downarrow\uparrow\downarrow\rangle_{A_{1}B_{1}C_{1}}|\downarrow\uparrow\downarrow\rangle_{A_{2}B_{2}C_{2}},
\end{eqnarray}
\begin{eqnarray}
|\phi_{3}\rangle &=&
\frac{1}{\sqrt{2}}(|\uparrow\uparrow\downarrow\rangle_{A_{1}B_{1}C_{1}}|\uparrow\uparrow\downarrow\rangle_{A_{2}B_{2}C_{2}}
+
|\downarrow\downarrow\uparrow\rangle_{A_{1}B_{1}C_{1}}|\downarrow\downarrow\uparrow\rangle_{A_{2}B_{2}C_{2}}.
\end{eqnarray}
When the parties all get the odd parity, the state should be
$\rho_{t_{odd}}$ (without normalization),
\begin{eqnarray}
\rho_{t_{odd}} &=&
\frac{1}{2}(F_{0}^{2}|\psi_{0}\rangle\langle\psi_{0}|+F_{1}^{2}|\psi_{1}\rangle\langle\psi_{1}|
+ F_{2}^{2}|\psi_{2}\rangle\langle\psi_{2}|
+F_{3}^{2}|\psi_{3}\rangle\langle\psi_{3}|).\label{ensemblerho2}
\end{eqnarray}
where
\begin{eqnarray}
|\psi_{0}\rangle &=&
\frac{1}{\sqrt{2}}(|\uparrow\uparrow\uparrow\rangle_{A_{1}B_{1}C_{1}}|\downarrow\downarrow\downarrow\rangle_{A_{2}B_{2}C_{2}}
+
|\downarrow\downarrow\downarrow\rangle_{A_{1}B_{1}C_{1}}|\uparrow\uparrow\uparrow\rangle_{A_{2}B_{2}C_{2}},
\end{eqnarray}
\begin{eqnarray}
|\psi_{1}\rangle &=&
\frac{1}{\sqrt{2}}(|\downarrow\uparrow\uparrow\rangle_{A_{1}B_{1}C_{1}}|\uparrow\downarrow\downarrow\rangle_{A_{2}B_{2}C_{2}}
+
|\uparrow\downarrow\downarrow\rangle_{A_{1}B_{1}C_{1}}|\downarrow\uparrow\uparrow\rangle_{A_{2}B_{2}C_{2}},
\end{eqnarray}
\begin{eqnarray}
|\psi_{2}\rangle &=&
\frac{1}{\sqrt{2}}(\uparrow\downarrow\uparrow\rangle_{A_{1}B_{1}C_{1}}|\downarrow\uparrow\downarrow\rangle_{A_{2}B_{2}C_{2}}
+
|\downarrow\uparrow\downarrow\rangle_{A_{1}B_{1}C_{1}}|\uparrow\downarrow\uparrow\rangle_{A_{2}B_{2}C_{2}},
\end{eqnarray}
\begin{eqnarray}
|\psi_{3}\rangle &=&
\frac{1}{\sqrt{2}}(|\uparrow\uparrow\downarrow\rangle_{A_{1}B_{1}C_{1}}|\downarrow\downarrow\uparrow\rangle_{A_{2}B_{2}C_{2}}
+
|\downarrow\downarrow\uparrow\rangle_{A_{1}B_{1}C_{1}}|\uparrow\uparrow\downarrow\rangle_{A_{2}B_{2}C_{2}}.
\end{eqnarray}
It is obvious that  Alice, Bob, and Charlie obtain the same outcomes
as the case in which they all obtain an even parity, after they
perform a bit-flip operation on each of the three electrons
$A_2B_2C_2$.  That is, they can obtain the result that the
complicated system composed of the six electrons
$A_{1}B_{1}C_{1}A_{2}B_{2}C_{2}$ is in a mixture state
$\rho'_{t_{even}}$ (without normalization)
\begin{eqnarray}
\rho'_{t_{even}} &=&
F_{0}^{2}|\phi_{0}\rangle\langle\phi_{0}|+F_{1}^{2}|\phi_{1}\rangle\langle\phi_{1}|
+ F_{2}^{2}|\phi_{2}\rangle\langle\phi_{2}|
+F_{3}^{2}|\phi_{3}\rangle\langle\phi_{3}|.\label{ensemblerho3}
\end{eqnarray}
We need  only discuss the case that the system is in the states
$\vert \phi_i\rangle$ with the probabilities $F_i^2$  below.

In the three down-spatial modes, a Hadamard ($H$) operation is
performed on each of the electrons $A_{2}$, $B_{2}$, and $C_{2}$,
which will lead to the transformation
\begin{eqnarray}
|\uparrow\rangle & \rightarrow &
\frac{1}{\sqrt{2}}(|\uparrow\rangle+|\downarrow\rangle),\;\;\;\;\;\;\;\;\;\;\;
|\downarrow\rangle   \rightarrow
\frac{1}{\sqrt{2}}(|\uparrow\rangle-|\downarrow\rangle).
\end{eqnarray}
That is,  the states $|\phi_{i}\rangle$ and $|\psi_{j}\rangle$
$(i,j=0,1,2,3 )$ will be changed into some other states. Alice, Bob,
and Charlie measure the spin states of  the down-spatial modes
$A_{2}B_{2}C_{2}$. It is not difficult to find that the outcomes can
be divided into two groups, by considering the number of the
spin-down electrons $N_{\downarrow}$. In the first case where
$N_{\downarrow}$ is even, that is, if they obtain the outcomes of
measurements on their electrons through the down-spatial modes
$|\uparrow\uparrow\uparrow\rangle_{A_{2}B_{2}C_{2}}$,
$|\downarrow\downarrow\uparrow\rangle_{A_{2}B_{2}C_{2}}$,
$|\uparrow\downarrow\downarrow\rangle_{A_{2}B_{2}C_{2}}$ or
$|\downarrow\uparrow\downarrow\rangle_{A_{2}B_{2}C_{2}}$, Alice,
Bob, and Charlie will get the three-electron GHZ states
$|\Phi_{0}^{+}\rangle=\frac{1}{\sqrt{2}}(|\uparrow\uparrow\uparrow\rangle+|\downarrow\downarrow\downarrow\rangle)_{A_{1}B_{1}C_{1}}$,
$|\Phi_{1}^{+}\rangle=\frac{1}{\sqrt{2}}(|\downarrow\uparrow\uparrow\rangle+|\uparrow\downarrow\downarrow\rangle)_{A_{1}B_{1}C_{1}}$,
$|\Phi_{2}^{+}\rangle=\frac{1}{\sqrt{2}}(|\uparrow\downarrow\uparrow\rangle+|\downarrow\uparrow\downarrow\rangle)_{A_{1}B_{1}C_{1}}$,
or
$|\Phi_{3}^{+}\rangle=\frac{1}{\sqrt{2}}(|\uparrow\uparrow\downarrow\rangle+|\downarrow\downarrow\uparrow\rangle)_{A_{1}B_{1}C_{1}}$
with the probabilities of $\frac{1}{2}F_{0}^{2}$,
$\frac{1}{2}F_{1}^{2}$, $\frac{1}{2}F_{2}^{2}$, or
$\frac{1}{2}F_{3}^{2}$, respectively. In the other case where
$N_{\downarrow}$ is odd,  that is, if they obtain the outcomes
$|\uparrow\uparrow\downarrow\rangle_{A_{2}B_{2}C_{2}}$,
$|\uparrow\downarrow\uparrow\rangle_{A_{2}B_{2}C_{2}}$,
$|\downarrow\uparrow\uparrow\rangle_{A_{2}B_{2}C_{2}}$ or
$|\downarrow\downarrow\downarrow\rangle_{A_{2}B_{2}C_{2}}$, they
will get the other three-electron GHZ states
$|\Phi_{0}^{-}\rangle=\frac{1}{\sqrt{2}}(|\uparrow\uparrow\uparrow\rangle-|\downarrow\downarrow\downarrow\rangle)_{A_{1}B_{1}C_{1}}$,
$|\Phi_{1}^{-}\rangle=\frac{1}{\sqrt{2}}(|\downarrow\uparrow\uparrow\rangle-|\uparrow\downarrow\downarrow\rangle)_{A_{1}B_{1}C_{1}}$,
$|\Phi_{2}^{-}\rangle=\frac{1}{\sqrt{2}}(|\uparrow\downarrow\uparrow\rangle-|\downarrow\uparrow\downarrow\rangle)_{A_{1}B_{1}C_{1}}$,
or
$|\Phi_{3}^{-}\rangle=\frac{1}{\sqrt{2}}(|\uparrow\uparrow\downarrow\rangle-|\downarrow\downarrow\uparrow\rangle)_{A_{1}B_{1}C_{1}}$
with the probabilities of $\frac{1}{2}F_{0}^{2}$,
$\frac{1}{2}F_{1}^{2}$, $\frac{1}{2}F_{2}^{2}$, or
$\frac{1}{2}F_{3}^{2}$, respectively. For the second case, in order
to obtain the GHZ state without phase-flip errors, the three
participants  should flip the relative phase of their electron
system $A_{1}B_{1}C_{1}$. For simplicity, here we  transform the
state $\rho_{t_{odd}}$ into $\rho_{t_{even}}$. In fact, we can
easily demonstrate that the conclusion is also suitable for
$\rho_{t_{odd}}$. In other words, the transformation is unnecessary.

Up to now, by keeping the instances in which all the parties obtain
the same parity,  and then measuring the electron spins from the
down-spatial modes after the H operations, the quantum state of the
three-electron system $A_1B_1C_1$ becomes $\rho'$. Here
\begin{eqnarray}
\rho'&=&
F_{0}^{'}|\Phi_{0}^{+}\rangle\langle\Phi_{0}^{+}|+F_{1}^{'}|\Phi_{1}^{+}\rangle\langle\Phi_{1}^{+}|
+
F_{2}^{'}|\Phi_{2}^{+}\rangle\langle\Phi_{2}^{+}|+F_{3}^{'}|\Phi_{3}^{+}\rangle\langle\Phi_{3}^{+}|.\label{ensemblerho}
\end{eqnarray}
where
\begin{eqnarray}
F_{0}^{'} &=&
\frac{F_{0}^{2}}{F_{0}^{2}+F_{1}^{2}+F_{2}^{2}+(1-F_{0}-F_{1}-F_{2})^{2}},
\nonumber \\
F_{1}^{'} &=&
\frac{F_{1}^{2}}{F_{0}^{2}+F_{1}^{2}+F_{2}^{2}+(1-F_{0}-F_{1}-F_{2})^{2}},
\nonumber \\
F_{2}^{'} &=&
\frac{F_{2}^{2}}{F_{0}^{2}+F_{1}^{2}+F_{2}^{2}+(1-F_{0}-F_{1}-F_{2})^{2}},
\nonumber\\
 F_{3}^{'}&=&\frac{(1-F_{0}-F_{1}-F_{2})^{2}}{F_{0}^{2}+F_{1}^{2}+F_{2}^{2}+(1-F_{0}-F_{1}-F_{2})^{2}}.
\end{eqnarray}
The fidelity of the new ensemble $F_{0}^{'}>F_{0}$ when the initial
fidelity $F_{0}$ satisfies the relation
\begin{eqnarray}
F_{0}&>&\frac{1}{4}
\left\{3-2F_{1}-2F_{2}-\sqrt{1+4(F_{1}+F_{2})-12(F_{1}^{2}+F_{2}^{2})-8F_{1}F_{2}}\right\}.\nonumber\\
 \end{eqnarray}
For more distinct, we take the case $F_{1}=F_{2}=F_{3}$ (a symmetric
noise model) as an example, and find that the fidelity of the state
$|\Phi_{0}^{+}\rangle$ will be improved by our normal MEPP just when
$F_{0}>\frac{1}{4}$. That is, the initial fidelity before EPP is
required to be $F_0
> \frac{1}{4}$, not the case in other MEPPs \cite{Murao,Horodecki,Yong,shengepjd,shengpla} in which it is required
to be  $F_0 > \frac{1}{2}$.

We have  fully discussed  our normal MEPP for  general bit-flip
errors in three-electron systems. We  use the PCDs based on charge
detection, instead of the perfect CNOT gates \cite{Murao}, to
fulfill the purification of bit-flip errors. Moreover, we give a
general form for the purification of bit-flip errors in
three-electron systems, not a Werner-type state \cite{Murao} or a
simplified mixed entangled state \cite{shengpla}, which makes our
normal MEPP have a higher efficiency than others
\cite{Murao,shengpla}. Especially, it doubles the efficiency of the
MEPP for three-electron systems in Ref.\cite{shengpla} as the three
parties not only consider the case in which they all obtain an even
parity but also the case they all obtain an odd parity.

\subsection{Recycling three-electron entanglement purification for bit-flip errors from subsystems}

In our normal three-electron EPP for bit-flip errors, the three
parties do not take the cross-combination items
$|\Phi_{i}^{+}\rangle_{A_{1}B_{1}C_{1}}\otimes|\Phi_{j}^{+}\rangle_{A_{2}B_{2}C_{2}}$
$(i\neq j \in\{0,1,2,3\})$ into account for obtaining high-fidelity
three-electron systems because they  obtain different parities. This
is just the flaw in all existing CEPPs
\cite{Bennett1,Deutsch,Pan1,Simon,shengpra,wangcpra,wangcqic,feng,Murao,Horodecki,Yong,shengepjd,shengpla}.
For these cross-combination items, when the three participants
perform some operations on their  electrons $A_{2}$, $B_{2}$ and
$C_{2}$, respectively, they cannot determine the state of the
remaining three electrons $A_{1}B_{1}C_{1}$ from the up-spatial
modes because the item
$|\Phi_{j}^{+}\rangle_{A_{1}B_{1}C_{1}}\otimes|\Phi_{i}^{+}\rangle_{A_{2}B_{2}C_{2}}$
has the same probability $F_{i}F_{j}$ as the item
$|\Phi_{i}^{+}\rangle
_{A_{1}B_{1}C_{1}}\otimes|\Phi_{j}^{+}\rangle_{A_{2}B_{2}C_{2}}$
$(i\neq j\in\{0,1,2,3\})$. That is, Alice, Bob, and Charlie will
obtain the state $|\Phi_{i}^{+}\rangle _{A_{1}B_{1}C_{1}}$ and
$|\Phi_{j}^{+}\rangle _{A_{1}B_{1}C_{1}}$ with the same probability.
These instances will decrease the fidelity of the state
$|\Phi_{0}^{+}\rangle _{A_{1}B_{1}C_{1}}$ in the three-electron
systems kept. This is just the reason that all existing CEPPs
discard  the cross-combination items. However, we cannot come to a
simple conclusion that the cross-combination items are useless,
because they can be used to distill some high-fidelity two-electron
entangled states. With a set of high-fidelity two-electron entangled
subsystems, Alice, Bob, and Charlie can produce a subset of
high-fidelity three-electron entangled systems with entanglement
link based on charge detection, which is far different from the
existing CEPPs
\cite{Bennett1,Deutsch,Pan1,Simon,shengpra,wangcpra,wangcqic,feng,Murao,Horodecki,Yong,shengepjd,shengpla},
including the MEPPs. We call this part of our MEPP  the recycling
MEPP. Our recycling MEPP will increase the efficiency and the yield
of our three-electron MEPP largely, especially in the case that the
original fidelity of the state $\vert \Phi^+_0\rangle_{ABC}$ is not
large.

In detail, our recycling EPP for three-electron systems includes
three steps. One is to distill a set of high-fidelity entangled
two-electron systems from the cross-combination items
$|\Phi_{i}^{+}\rangle_{A_{1}B_{1}C_{1}}\otimes|\Phi_{j}^{+}\rangle_{A_{2}B_{2}C_{2}}$
$(i\neq j \in\{0,1,2,3\})$. The second step is to improve the
fidelity of subsystems with a two-electron EPP. The third step is to
produce entangled three-electron systems from subspaces with
entanglement link based on a PCD.

\subsubsection{Two-electron entanglement distillation from the cross-combination items of three-electron systems}

We take the two cross-combination items
$|\varphi\rangle_{1}=|\Phi_{0}^{+}\rangle_{A_{1}B_{1}C_{1}}\otimes|\Phi_{1}^{+}\rangle_{A_{2}B_{2}C_{2}}$
and
$|\varphi\rangle_{2}=|\Phi_{1}^{+}\rangle_{A_{1}B_{1}C_{1}}\otimes|\Phi_{0}^{+}\rangle_{A_{2}B_{2}C_{2}}$
as an example to demonstrate the principle of our two-electron
entanglement distillation from three-electron systems with bit-flip
errors. As for the other cross-combination items, we could deal with
them in the same way with or without a little modification. It is
interesting to point out that whether the cross-combination items is
$|\varphi\rangle_{1}$ or $|\varphi\rangle_{2}$, the three parties
will obtain the maximally entangled two-electron state
$|\phi^{+}\rangle_{B_{1}C_{1}} \equiv
\frac{1}{\sqrt{2}}(|\uparrow\uparrow\rangle +
|\downarrow\downarrow\rangle)_{B_{1}C_{1}}$, by measuring the
electron spins with potential errors.

To write the states $|\varphi\rangle_{1}$ and $|\varphi\rangle_{2}$
in a detail way, they can be described as
\begin{eqnarray}
|\varphi\rangle_{1} &=&
\frac{1}{2}(|\uparrow\uparrow\uparrow\rangle_{A_{1}B_{1}C_{1}}|\uparrow\downarrow\downarrow\rangle_{A_{2}B_{2}C_{2}}+
|\downarrow\downarrow\downarrow\rangle_{A_{1}B_{1}C_{1}}|\downarrow\uparrow\uparrow\rangle_{A_{2}B_{2}C_{2}}\nonumber\\
&+&
|\uparrow\uparrow\uparrow\rangle_{A_{1}B_{1}C_{1}}|\downarrow\uparrow\uparrow\rangle_{A_{2}B_{2}C_{2}}+
|\downarrow\downarrow\downarrow\rangle_{A_{1}B_{1}C_{1}}|\uparrow\downarrow\downarrow\rangle_{A_{2}B_{2}C_{2}}),
\end{eqnarray}
\begin{eqnarray}
|\varphi\rangle_{2} &=&
\frac{1}{2}(|\uparrow\downarrow\downarrow\rangle_{A_{1}B_{1}C_{1}}|\uparrow\uparrow\uparrow\rangle_{A_{2}B_{2}C_{2}}+
|\downarrow\uparrow\uparrow\rangle_{A_{1}B_{1}C_{1}}|\downarrow\downarrow\downarrow\rangle_{A_{2}B_{2}C_{2}}\nonumber\\
&+&
|\downarrow\uparrow\uparrow\rangle_{A_{1}B_{1}C_{1}}|\uparrow\uparrow\uparrow\rangle_{A_{2}B_{2}C_{2}}+
|\uparrow\downarrow\downarrow\rangle_{A_{1}B_{1}C_{1}}|\downarrow\downarrow\downarrow\rangle_{A_{2}B_{2}C_{2}}).
\end{eqnarray}
One can see that if the outcomes of parity-check measurements done
by Alice, Bob and Charlie are even, odd, and odd, respectively, the
six-electron system is in the state
\begin{eqnarray}
|\zeta\rangle_{1} &\equiv&
\frac{1}{\sqrt{2}}(|\uparrow\uparrow\uparrow\rangle_{A_{1}B_{1}C_{1}}
|\uparrow\downarrow\downarrow\rangle_{A_{2}B_{2}C_{2}}+
|\downarrow\downarrow\downarrow\rangle_{A_{1}B_{1}C_{1}}|\downarrow\uparrow\uparrow\rangle_{A_{2}B_{2}C_{2}})
\end{eqnarray}
which comes from the state $|\varphi\rangle_{1}$, or
\begin{eqnarray}
|\zeta\rangle_{2} &\equiv&
\frac{1}{\sqrt{2}}(|\uparrow\downarrow\downarrow\rangle_{A_{1}B_{1}C_{1}}
|\uparrow\uparrow\uparrow\rangle_{A_{2}B_{2}C_{2}}+
|\downarrow\uparrow\uparrow\rangle_{A_{1}B_{1}C_{1}}|\downarrow\downarrow\downarrow\rangle_{A_{2}B_{2}C_{2}})
\end{eqnarray}
from $|\varphi\rangle_{2}$ with the same probability of
$\frac{1}{2}F_{0}F_{1}$. If the outcomes are odd, even, and even,
the whole state of the system is
\begin{eqnarray}
|\zeta\rangle_{3} &\equiv&
\frac{1}{\sqrt{2}}(|\uparrow\uparrow\uparrow\rangle_{A_{1}B_{1}C_{1}}|\downarrow\uparrow\uparrow\rangle_{A_{2}B_{2}C_{2}}
+
|\downarrow\downarrow\downarrow\rangle_{A_{1}B_{1}C_{1}}|\uparrow\downarrow\downarrow\rangle_{A_{2}B_{2}C_{2}})
\end{eqnarray}
from $|\varphi\rangle_{1}$, or
\begin{eqnarray}
|\zeta\rangle_{4} &\equiv&
\frac{1}{\sqrt{2}}(|\downarrow\uparrow\uparrow\rangle_{A_{1}B_{1}C_{1}}|\uparrow\uparrow\uparrow\rangle_{A_{2}B_{2}C_{2}}+
|\uparrow\downarrow\downarrow\rangle_{A_{1}B_{1}C_{1}}|\downarrow\downarrow\downarrow\rangle_{A_{2}B_{2}C_{2}})
\end{eqnarray}
from $|\varphi\rangle_{2}$ with the same probability of
$\frac{1}{2}F_{0}F_{1}$ too. With an $H$ operation  on each of the
four electrons $A_1$, $A_2$, $B_2$, and $C_2$, the state
$|\zeta\rangle_{1}$ will be transformed into
\begin{eqnarray}
|\zeta\rangle_{1}^{H} &=& \frac{1}{2} \{(|\uparrow\uparrow\rangle +
|\downarrow\downarrow\rangle)_{B_{1}C_{1}}(|\uparrow\uparrow\uparrow\uparrow\rangle
+
|\downarrow\downarrow\uparrow\uparrow\rangle+|\uparrow\uparrow\downarrow\downarrow\rangle
+|\downarrow\downarrow\downarrow\downarrow\rangle -
|\downarrow\uparrow\downarrow\uparrow\rangle-|\uparrow\downarrow\downarrow\uparrow\rangle-|\downarrow\uparrow\uparrow\downarrow\rangle
\nonumber
\\
&-& |\uparrow\downarrow\uparrow\downarrow\rangle)_{A_1A_2B_2C_2} +
(|\uparrow\uparrow\rangle-|\downarrow\downarrow\rangle)_{B_{1}C_{1}}(|\downarrow\uparrow\uparrow\uparrow\rangle
+
|\uparrow\downarrow\uparrow\uparrow\rangle+|\uparrow\downarrow\downarrow\downarrow\rangle
+|\downarrow\downarrow\downarrow\uparrow\rangle
-|\uparrow\downarrow\downarrow\uparrow\rangle
\nonumber\\
&-&
|\downarrow\downarrow\uparrow\downarrow\rangle-|\downarrow\uparrow\uparrow\downarrow\rangle
-|\uparrow\uparrow\downarrow\uparrow\rangle)_{A_1A_2B_2C_2}\}.\nonumber\\
\end{eqnarray}
In order to distill a two-electron entangled state, Bob and Charlie
detect their electrons $B_{2}$ and $C_{2}$, and Alice detects her
two electrons $A_{1}$ and $A_{2}$ with the basis
$\sigma_z\equiv\{|\uparrow\rangle,|\downarrow\rangle\}$,
respectively. When the occupation number of $|\downarrow\rangle$ in
the outcomes is even (such as
$|\uparrow\uparrow\uparrow\uparrow\rangle$,
$|\downarrow\downarrow\uparrow\uparrow\rangle$,
$|\downarrow\uparrow\uparrow\downarrow\rangle$, and so on), Bob and
Charlie obtain the two-electron  entangled state
$|\phi^{+}\rangle_{B_{1}C_{1}}$ with the probability of
$\frac{1}{4}F_{0}F_{1}$. When it is odd, they can obtain the state
$|\phi^{-}\rangle_{B_{1}C_{1}}\equiv\frac{1}{\sqrt{2}}(|\uparrow\uparrow\rangle-|\downarrow\downarrow\rangle)$
with the same probability of $\frac{1}{4}F_{0}F_{1}$. The state
$|\phi^{-}\rangle_{B_{1}C_{1}}$ can be transformed into the state
$|\phi^{+}\rangle$  by flipping its relative phase. As for the cases
$|\zeta\rangle_{2}$,  $|\zeta\rangle_{3}$, and $|\zeta\rangle_{4}$,
the same conclusions can be drawn  by simple calculations. Thus, the
total probability of obtaining $|\phi^{+}\rangle_{B_{1}C_{1}}$ from
the cross-combination items $|\varphi\rangle_{1}$ and
$|\varphi\rangle_{2}$ is $2F_{0}F_{1}$.

Up to now, the principle of distilling two-electron entangled states
from the cross-combination items
$|\Phi_{0}^{+}\rangle_{A_{1}B_{1}C_{1}}\otimes|\Phi_{1}^{+}\rangle_{A_{2}B_{2}C_{2}}$
and
$|\Phi_{1}^{+}\rangle_{A_{1}B_{1}C_{1}}\otimes|\Phi_{0}^{+}\rangle_{A_{2}B_{2}C_{2}}$
in which only one bit-flip error takes place on  Alice's electrons,
has been completely discussed. As for the other cross-combination
items with one bit-flip error, using the same process described
above, the three parties can obtain the two-electron entangled
states $|\phi^{+}\rangle_{A_{1}C_{1}} \equiv
\frac{1}{\sqrt{2}}(|\uparrow\uparrow\rangle+|\downarrow\downarrow\rangle)_{A_{1}C_{1}}$
and $|\phi^{+}\rangle_{A_{1}B_{1}} \equiv
\frac{1}{\sqrt{2}}(|\uparrow\uparrow\rangle+|\downarrow\downarrow\rangle)_{A_{1}B_{1}}$
with the probabilities of $2F_{0}F_{2}$ and $2F_{0}F_{3}$,
respectively. When we come to the cases that there is  a bit-flip
error on both the systems $A_{1}B_{1}C_{1}$ and $A_{2}B_{2}C_{2}$
(the two electrons with a bit-flip error belong to different
participants), that is,
$|\Phi_{i}^{+}\rangle_{A_{1}B_{1}C_{1}}\otimes|\Phi_{j}^{+}\rangle_{A_{2}B_{2}C_{2}}(i\neq
j\in\{1,2,3\})$, Alice, Bob and Charlie  can also deal with them in
the same way. However, the two-electron states obtained will be the
ones with bit-flip errors
$|\psi^{+}\rangle_{A_{1}C_{1}}\equiv\frac{1}{\sqrt{2}}(|\uparrow\downarrow\rangle+|\downarrow\uparrow\rangle)_{A_{1}C_{1}}$,
$|\psi^{+}\rangle_{A_{1}B_{1}}\equiv\frac{1}{\sqrt{2}}(|\uparrow\downarrow\rangle+|\downarrow\uparrow\rangle)_{A_{1}B_{1}}$,
and
$|\psi^{+}\rangle_{B_{1}C_{1}}\equiv\frac{1}{\sqrt{2}}(|\uparrow\downarrow\rangle+|\downarrow\uparrow\rangle)_{B_{1}C_{1}}$
with the probabilities $2F_{1}F_{3}$, $2F_{1}F_{2}$, and
$2F_{2}F_{3}$, respectively. That is, the states of the two-electron
systems kept can be described as (without normalization)
\begin{eqnarray}
\rho_{_{AB}}=2F_{0}F_{3}|\phi^{+}\rangle_{AB}\langle\phi^{+}|+2F_{1}F_{2}|\psi^{+}\rangle_{AB}\langle\psi^{+}|,\nonumber \\
\rho_{_{AC}}=2F_{0}F_{2}|\phi^{+}\rangle_{AC}\langle\phi^{+}|+2F_{1}F_{3}|\psi^{+}\rangle_{AC}\langle\psi^{+}|,\nonumber \\
\rho_{_{BC}}=2F_{0}F_{1}|\phi^{+}\rangle_{BC}\langle\phi^{+}|+2F_{2}F_{3}|\psi^{+}\rangle_{BC}\langle\psi^{+}|.\label{dm2e}
\end{eqnarray}
The fidelity $F^b_i \equiv \frac{F_{0}F_{i}}{F_{0}F_{i}+F_{j}F_{k}}$
($i$, $j$ and $k$  are  different  from  each  other, $i, j, k
\in\{1,2,3\}$) of the two-electron subsystems  in the state $\vert
\phi^+\rangle$ is larger than that of the initial three-electron
systems $F_0$ when the relation $F_{0}<1-\frac{F_{j}F_{k}}{F_{i}}$
is satisfied. Let us take the symmetric noise model
$F_{1}=F_{2}=F_{3}$ and $F_{0}>F_{1}$ as an example to show the
relation between $F^b_i$ and $F_0$. The inequality equation can be
simplified to be $F_{0}+F_{1}<1$ and the fidelity of the distilled
two-electron subsystems $\frac{F_{0}}{F_{0}+F_{1}}$ is
unconditionally larger than that of the transmitted three-electron
systems $F_0$. That is to say, the parties can distill a
two-electron spin subsystems with a fidelity higher than  $F_0$ from
each cross-combination item discarded in all conventional MEPPs.

\subsubsection{Two-electron entanglement purification based on charge detection}

After obtaining a set of two-electron entangled subsystems, Alice,
Bob, and Charlie can first improve the fidelity of their
two-electron ensembles and then produce a subset of three-electron
systems with entanglement link based on charge detection. The
principle of our EPP for two-electron subsystems is  similar to all
existing conventional two-qubit EPPs
\cite{Bennett1,Deutsch,Pan1,Simon,shengpra,wangcpra,wangcqic,feng}.
Let us use the two parties Alice and Bob to describe its principle
clearly. The principle of two-electron EPP for any other two parties
is the same as this one.

Suppose that the two-electron ensemble obtained by Alice and Bob is
in the state
\begin{eqnarray}
\rho^{AB}_{0} = f_{0}|\phi^+\rangle_{AB}\langle\phi^+| +
f_{1}|\psi^+\rangle_{AB}\langle\psi^+|.\label{two-electronensemblerho0}
\end{eqnarray}
For each pair of two-electron subsystems, say $A_1B_1$ and $A_2B_2$,
their state is $\rho^{AB}_0\otimes \rho^{AB}_0$. It can be viewed as
the mixture of the 4 pure entangled states, that is,
$|\phi^+\rangle_{A_1B_1} \otimes |\phi^+\rangle_{A_2B_2}$,
$|\phi^+\rangle_{A_1B_1} \otimes |\psi^+\rangle_{A_2B_2}$,
$|\psi^+\rangle_{A_1B_1} \otimes |\phi^+\rangle_{A_2B_2}$, and
$|\psi^+\rangle_{A_1B_1} \otimes |\psi^+\rangle_{A_2B_2}$ with the
probabilities of $f_0f_0$, $f_0f_1$, $f_1f_0$, and  $f_1f_1$,
respectively. Alice  let her two electrons $A_1$ and $A_2$ pass
through her PCD, shown in Fig.\ref{fig2_MEEPP}. So does Bob. Alice
and Bob keep the instances in which they both obtain an even parity
or an odd parity.

When both Alice and Bob obtain an even parity, the four-electron
system $A_1B_1A_2B_2$ is in the states
\begin{eqnarray}
|\lambda_1 \rangle = \frac{1}{\sqrt{2}} (| \uparrow \uparrow
\rangle_{A_1B_1}| \uparrow \uparrow \rangle_{A_2B_2} + | \downarrow
 \downarrow \rangle_{A_1B_1}|  \downarrow \downarrow
 \rangle_{A_2B_2})\nonumber
\end{eqnarray}
and
\begin{eqnarray}
|\lambda_2 \rangle = \frac{1}{\sqrt{2}} (| \uparrow \downarrow
\rangle_{A_1B_1}| \uparrow \downarrow \rangle_{A_2B_2} + |
\downarrow
 \uparrow \rangle_{A_1B_1}|  \downarrow \uparrow \rangle_{A_2B_2})
\nonumber
\end{eqnarray}
with the probabilities of $\frac{1}{2} f_0f_0$ and $\frac{1}{2}
f_1f_1$, respectively. When they both obtain an odd parity, the
four-electron system  is in the states
\begin{eqnarray}
|\lambda_3 \rangle = \frac{1}{\sqrt{2}} (| \uparrow \uparrow
\rangle_{A_1B_1}| \downarrow \downarrow \rangle_{A_2B_2} + |
\downarrow
 \downarrow \rangle_{A_1B_1}|  \uparrow \uparrow \rangle_{A_2B_2})
\nonumber
\end{eqnarray}
and
\begin{eqnarray}
|\lambda_4 \rangle = \frac{1}{\sqrt{2}} (| \uparrow \downarrow
\rangle_{A_1B_1}| \downarrow \uparrow \rangle_{A_2B_2} + |
\downarrow
 \uparrow \rangle_{A_1B_1}|  \uparrow \downarrow \rangle_{A_2B_2})
\nonumber
\end{eqnarray}
with the probabilities of $\frac{1}{2} f_0f_0$ and $\frac{1}{2}
f_1f_1$, respectively. With a bit-flip operation $\sigma_x =
|\uparrow\rangle\langle \downarrow| + |\downarrow\rangle\langle
\uparrow| $ on each of the two electrons $A_2B_2$, the states
$|\lambda_3 \rangle$ and $|\lambda_4 \rangle$ are transformed into
the states $|\lambda_1 \rangle$ and $|\lambda_2 \rangle$,
respectively. That is, Alice and Bob can obtain the states
$|\lambda_1 \rangle$ and $|\lambda_2 \rangle$ with the probabilities
of $f_0f_0$ and $f_1f_1$, respectively, when they obtain the same
parity.

After an $H$ operation on each of the two electrons $A_2$ and $B_2$,
Alice and Bob measure the spins of these two electrons with the
basis $\sigma_z$. If they obtain the same spin states (that is, both
the spin-up state or the spin-down state), the two-electron
subsystem $A_1B_1$ kept by Alice and Bob is in the states $\vert
\phi^+\rangle_{A_1B_1}$ and $\vert \psi^+\rangle_{A_1B_1}$ with the
probabilities of $\frac{1}{2} f_0f_0$ and $\frac{1}{2} f_1f_1$,
respectively. If they obtain two different spin states (that is, one
is the spin-up state and the other is the spin-down state), the
two-electron subsystem $A_1B_1$ is in the states $\vert
\phi^-\rangle_{A_1B_1}$ and $\vert \psi^-\rangle_{A_1B_1}$ with the
probabilities of $\frac{1}{2} f_0f_0$ and $\frac{1}{2} f_1f_1$,
respectively. In this time, Alice and Bob can completely transform
the states $\vert \phi^-\rangle_{A_1B_1}$ and $\vert
\psi^-\rangle_{A_1B_1}$ into the states $\vert
\phi^+\rangle_{A_1B_1}$ and $\vert \psi^+\rangle_{A_1B_1}$,
respectively. That is to say, the two-electron ensemble after a
round of purification is in the state (without normalization)
\begin{eqnarray}
\rho^{AB}_{1} = f^2_{0}|\phi^+\rangle_{AB}\langle\phi^+| +
f^2_{1}|\psi^+\rangle_{AB}\langle\psi^+|.\label{two-electronensemblerho1}
\end{eqnarray}

After Alice and Bob perform $n$ times of this two-electron EPP, the
two-electron ensemble kept is in the state (without normalization)
\begin{eqnarray}
\rho^{AB}_{n} = f^{2^n}_{0}|\phi^+\rangle_{AB}\langle\phi^+| +
f^{2^n}_{1}|\psi^+\rangle_{AB}\langle\psi^+|.\label{two-electronensemblerhon}
\end{eqnarray}
The fidelity of the state $\vert \phi^+\rangle$ is
\begin{eqnarray}
f'_{n} = \frac{f^{2^n}_{0}}{f^{2^n}_{0} +
f^{2^n}_{1}}.\label{fidelityn}
\end{eqnarray}

\subsubsection{Three-electron entanglement production from two-electron
subsystems with entanglement link}

With a set of high-fidelity two-electron entangled subsystems,  the
three parties can create a subset of high-fidelity three-electron
entangled systems nonlocally with entanglement link. As an example,
we can use the case with three symmetric channels
$F_{1}=F_{2}=F_{3}=\frac{1-F_0}{3}$  to show the principle of
three-electron entanglement production from two-electron subsystems
below.

\begin{figure}[!h]
\begin{center}
\includegraphics[width=5.4cm,angle=0]{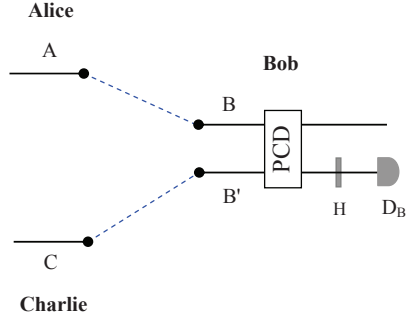}
\caption{The principle of the entanglement production of a
three-electron system from two two-electron entangled subsystems
with entanglement link based on a PCD.} \label{fig4_two_qubit}
\end{center}
\end{figure}

Without the two-electron entanglement purification process, the
density matrices in Eq.(\ref{dm2e}) are reduced to be (without
normalization)
\begin{eqnarray}
\rho_{_{AB}}^{s}=2F_{0}F_1|\phi^{+}\rangle_{AB}\langle\phi^{+}| + 2F_{1}^{2}|\psi^{+}\rangle_{AB}\langle\psi^{+}|,\nonumber \\
\rho_{_{AC}}^{s}=2F_{0}F_1|\phi^{+}\rangle_{AC}\langle\phi^{+}| + 2F_{1}^{2}|\psi^{+}\rangle_{AC}\langle\psi^{+}|,\nonumber \\
\rho_{_{BC}}^{s}=2F_{0}F_1|\phi^{+}\rangle_{BC}\langle\phi^{+}| +
2F_{1}^{2}|\psi^{+}\rangle_{BC}\langle\psi^{+}|. \label{rho2s}
\end{eqnarray}
After $n$ times of the two-electron EPP are performed, the fidelity
of the two-electron state $\vert\phi^+\rangle$ is improved largely
and the density matrices in Eq.(\ref{rho2s})  become (without
normalization)
\begin{eqnarray}
\rho_{_{AB}}^{sn}=(2F_{0}F_1)^{2^n} |\phi^{+}\rangle_{AB}\langle\phi^{+}| + (2F_{1}^{2})^{2^n} |\psi^{+}\rangle_{AB}\langle\psi^{+}|,\nonumber \\
\rho_{_{AC}}^{sn}=(2F_{0}F_1)^{2^n} |\phi^{+}\rangle_{AC}\langle\phi^{+}| + (2F_{1}^{2})^{2^n} |\psi^{+}\rangle_{AC}\langle\psi^{+}|,\nonumber \\
\rho_{_{BC}}^{sn}=(2F_{0}F_1)^{2^n}
|\phi^{+}\rangle_{BC}\langle\phi^{+}| +
(2F_{1}^{2})^{2^n}|\psi^{+}\rangle_{BC}\langle\psi^{+}|.\nonumber\\
\end{eqnarray}
Let us assume that $F_{0}^{s}\equiv (2F_{0}F_1)^{2^n}$ and
$F_{1}^{s}\equiv(2F_{1}^{2})^{2^n}$.

In principle, Alice, Bob, and Charlie can produce a three-electron
entangled system from two two-electron subsystems with a high
fidelity. Let us take the two-electron subsystems $AB$ and $B'C$ as
an example to describe the principle of entanglement production from
two-electron subsystems, shown in Fig.\ref{fig4_two_qubit}.  The
state of the complicated quantum system composed of four electrons
$A$, $B$, $B'$, and $C$ can be viewed as the mixture of four pure
states. That is, it is in the states
$|\phi^{+}\rangle_{AB}\otimes|\phi^{+}\rangle_{B^{'}C}$,
$|\phi^{+}\rangle_{AB}\otimes|\psi^{+}\rangle_{B^{'}C}$,
$|\psi^{+}\rangle_{AB}\otimes|\phi^{+}\rangle_{B^{'}C}$, and
$|\psi^{+}\rangle_{AB}\otimes|\psi^{+}\rangle_{B^{'}C}$ with the
probabilities of $F_{0}^{s}F_{0}^{s}$,  $F_{0}^{s}F_{1}^{s}$,
$F_{1}^{s}F_{0}^{s}$, and  $F_{1}^{s}F_{1}^{s}$, respectively. Bob
performs a parity-check detection on his two electrons $B$ and $B'$,
and then he divides the four-electron  system into two cases
according to the outcomes obtained, i.e., an even-parity case and an
odd-parity case. When Bob obtains the even parity, the four-electron
system is in the states
\begin{eqnarray}
\vert \Lambda_1\rangle &=&
\frac{1}{\sqrt{2}}(|\uparrow\uparrow\uparrow\uparrow\rangle_{ABB^{'}C}
+|\downarrow\downarrow\downarrow\downarrow\rangle_{ABB^{'}C}),\\
\vert \Lambda_2\rangle &=&
\frac{1}{\sqrt{2}}(|\uparrow\uparrow\uparrow\downarrow\rangle_{ABB^{'}C}
+|\downarrow\downarrow\downarrow\uparrow\rangle_{ABB^{'}C}),\\
\vert \Lambda_3\rangle &=&
\frac{1}{\sqrt{2}}(|\downarrow\uparrow\uparrow\uparrow\rangle_{ABB^{'}C}
+|\uparrow\downarrow\downarrow\downarrow\rangle_{ABB^{'}C}),
\end{eqnarray}
and
\begin{eqnarray}
\vert \Lambda_4\rangle &=&
\frac{1}{\sqrt{2}}(|\uparrow\downarrow\downarrow\uparrow\rangle_{ABB^{'}C}+|\downarrow\uparrow\uparrow\downarrow\rangle_{ABB^{'}C})
\end{eqnarray}
with the probabilities $\frac{1}{2}F_{0}^{s}F_{0}^{s}$,
$\frac{1}{2}F_{0}^{s}F_{1}^{s}$, $\frac{1}{2}F_{1}^{s}F_{0}^{s}$,
and $\frac{1}{2}F_{1}^{s}F_{1}^{s}$, respectively. To obtain the
three-electron entangled state, Bob  detects the spin of his
electron $B^{'}$ with the basis
$\sigma_z\equiv\{|\uparrow\rangle,|\downarrow\rangle\}$ after an $H$
operation is performed on it. When he obtains
$|\uparrow\rangle_{B'}$, the three electrons $ABC$ are in a mixed
state by mixing four pure states $|\Phi^{+}_{0}\rangle$,
$|\Phi^{+}_{3}\rangle$, $|\Phi^{+}_{1}\rangle$, and
$|\Phi^{+}_{2}\rangle$ with the probabilities of
$\frac{1}{4}F_{0}^{s}F_{0}^{s}$, $\frac{1}{4}F_{0}^{s}F_{1}^{s}$,
$\frac{1}{4}F_{1}^{s}F_{0}^{s}$, and
$\frac{1}{4}F_{1}^{s}F_{1}^{s}$, respectively. When  the state of
the electron $B^{'}$ is $|\downarrow\rangle_{B'}$, they can obtain
the same result with a phase-flip operation on the electron $A$.
Therefore, the total probabilities that Alice, Bob, and Charlie
obtain the states $|\Phi^{+}_{0}\rangle$, $|\Phi^{+}_{1}\rangle$,
$|\Phi^{+}_{2}\rangle$, and $|\Phi^{+}_{3}\rangle$   are doubled
eventually. As for the odd-parity case, the process utilized is
nearly the same as the even-parity case but with an additional
bit-flip operation $\sigma_x$ performed on electron $C$. That is,
the state of the three electrons $ABC$ produced with entanglement
link can be written as
\begin{eqnarray}
\rho_{T} &=& (F_{0}^{s})^2 |\Phi^{+}_{0}\rangle\langle\Phi^{+}_{0}|
+ F_{1}^{s}F_{0}^{s} |\Phi^{+}_{1}\rangle\langle\Phi^{+}_{1}|+
(F_{1}^{s})^2 |\Phi^{+}_{2}\rangle\langle\Phi^{+}_{2}| +
F_{0}^{s}F_{1}^{s}|\Phi^{+}_{3}\rangle\langle\Phi^{+}_{3}|\nonumber\\
&=& (2F_{0}F_1)^{2^{n+1}} |\Phi^{+}_{0}\rangle\langle\Phi^{+}_{0}| +
(2F_{0}F_1)^{2^{n}}(2F_1^2)^{2^n}
|\Phi^{+}_{1}\rangle\langle\Phi^{+}_{1}| +  (2F_1^2)^{2^{n+1}} |\Phi^{+}_{2}\rangle\langle\Phi^{+}_{2}| \nonumber\\
&+& (2F_{0}F_1)^{2^{n}}(2F_1^2)^{2^n}
|\Phi^{+}_{3}\rangle\langle\Phi^{+}_{3}|.
\end{eqnarray}
The fidelity of the three-electron state $\vert
\Phi^+_0\rangle_{ABC}$ is
\begin{eqnarray}
F_{0}^{T} &=& \frac{F_{0}^{2^{n+1}}}{(F_{0}^{2^{n}} + F_1^{2^n})^2}.
\end{eqnarray}
$F_{0}^{T}>F_{0}$ when $F_{0}>\frac{1}{4}$.

\subsection{Numerical comparisons of efficiency, fidelity, and yield between our MEPP and conventional MEPPs}

Usually, the efficiency of an EPP $E$ is defined as the probability
that the parties can obtain a high-fidelity entangled three-electron
system (or entangled two-electron subsystems)  from a pair of
low-fidelity systems transmitted over a noisy channel without loss
after the parties perform a round of the EPP. The yield of an EPP
$Y$ is defined as the probability that the parties can obtain an
entangled three-electron system with the fidelity higher than a
threshold value $F_{thr}$,  from a pair of low-fidelity systems
transmitted after some rounds of the EPP are performed. It is
obvious that the efficiency and the yield of our three-electron EPP
$E_o$ and $Y_o$ depends on the three parameters $F_1$, $F_2$, and
$F_3$. For simplicity, we only discuss the case with the parameters
$F_1=F_2=F_3=\frac{1-F_0}{3}$ below to show the difference between
our MEPP and conventional ones clearly. Let us assume that the
threshold value of the final fidelity after purification is
$F_{thr}=0.95$, which means that the parties should improve the
fidelity of the three-electron systems kept to be $F'\geq F_{thr}$
by repeating the MEPP some times. For comparison, we use the
efficiency obtained with our normal EPP  $E_n$ to represent that in
conventional MEPPs as it is just the maximal efficiency that the
parties can obtain from the identity-combination items.

\begin{figure}[!h]
\begin{center}
\includegraphics[width=8cm,angle=0]{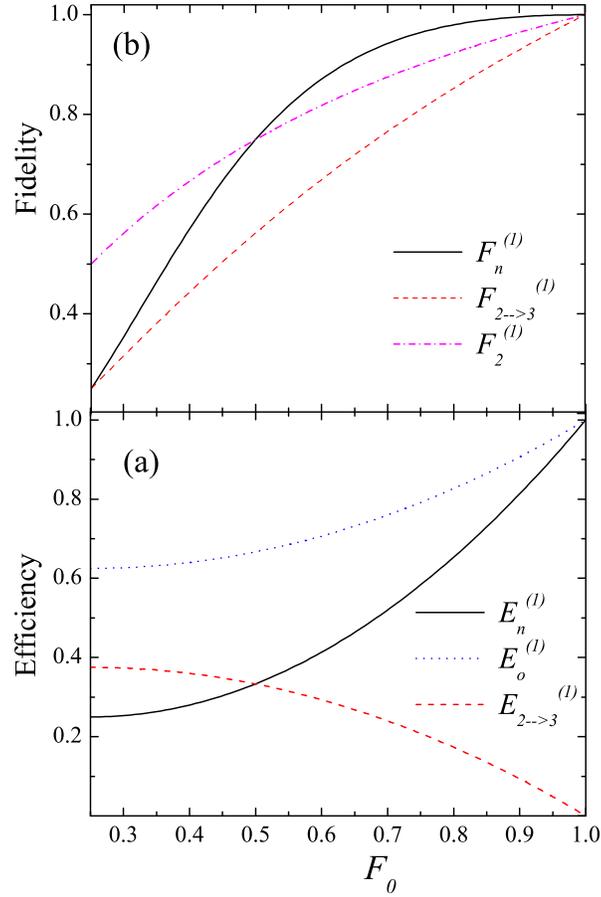}
\caption{Numerical comparison for the efficiency and fidelity
between our MEPP and others. (a) The efficiency of our EPP
$E^{(1)}_o$ and the maximal value of efficiency from the
conventional MEPPs $E^{(1)}_n$ for three-electron systems under a
symmetric noise ($F_1=F_2=F_3=\frac{1-F_0}{3}$) are shown with a
blue dot line and a black solid
 line, respectively. Here $E^{(1)}_{2\rightarrow 3}$ is the efficiency that
the three parties can obtain three-electron systems from
two-electron systems with entanglement link based on charge
detection. (b) The fidelity of the present MEPP $F_e$ and that of
the conventional MEPP $F_n$. Here $F_2$ and $F_{2\rightarrow 3}$ is
the fidelities of the two-photon systems obtained from the
cross-combinations and that of the three-photon systems obtained
directly from two-photon systems with entanglement link,
respectively. $F_0$ is just the original fidelity of three-photon
systems before entanglement purification.} \label{fig5_efficiency}
\end{center}
\end{figure}

By running our normal three-photon EPP only once, the efficiency of
the three-photon EPP $E_{n}^{(1)}$ is
\begin{eqnarray}
E_{n}^{(1)} &=& F^2_0+F^2_1+F^2_2+F^2_3 = \frac{1 - 2F_0 +
4F^2_0}{3}.
\end{eqnarray}
It is just the probability that the pair of three-electron systems
are in the identity-combination items $\vert
\Phi^+_i\rangle_{A_1B_1C_1} \otimes \vert
\Phi^+_i\rangle_{A_2B_2C_2}$ ($i=0,1,2,3$). After a round of
entanglement purification, the fidelity of the three-electron
systems kept (i.e., the relative probability of the state $\vert
\Phi^+_0\rangle_{ABC}$) becomes
\begin{eqnarray}
F_{n}^{(1)} &=& \frac{F_0^2}{F^2_0+F^2_1+F^2_2+F^2_3}  =
\frac{3F_0^2}{1 - 2F_0 + 4F^2_0}.
\end{eqnarray}

As each cross-combination item  $\vert\Phi_i^+\rangle \otimes
\vert\Phi_j^+\rangle$ ($i\neq j \in\{0,1,2,3\}$) will lead the three
parties to obtain an entangled two-electron subsystem, the
probability  $P^{(1)}_{3\rightarrow  2}$ that the three parties
obtain two-electron subsystems from a pair of three-electron systems
in the cross-combination items is
\begin{eqnarray}
P^{(1)}_{3\rightarrow  2} &=& \sum_{j\neq l=0}^3 F_jF_l = F_0(F_1 + F_2 + F_3) + F_1(F_0 + F_2 + F_3) \nonumber\\
&+& F_2(F_0 + F_1 + F_3) + F_3(F_0 + F_1 + F_2)\nonumber\\
&=& \frac{2 + 2F_0 -4F_0^2}{3}.
\end{eqnarray}
Because Alice, Bob, and Charlie can in principle obtain a
three-electron system from a pair of two-electron subsystems with
entanglement link based on charge detection if they do not improve
the fidelity of their two-electron subsystems before entanglement
link, the efficiency that the three parties obtain three-electron
entangled systems from two-electron entangled subsystems
$E^{(1)}_{2\rightarrow  3}$ is a half of $P^{(1)}_{3\rightarrow 2}$,
that is,
\begin{eqnarray}
E^{(1)}_{2\rightarrow  3} = \frac{1}{2}P^{(1)}_{3\rightarrow  2} =
\frac{1 + F_0 - 2F_0^2}{3}.
\end{eqnarray}

Taking three-electron entanglement production with entanglement link
into account, the efficiency of our MEPP $E^{(1)}_o$ for
three-electron systems after the three parties accomplish a round of
entanglement purification for bit-flip errors is
\begin{eqnarray}
E^{(1)}_o &=& E^{(1)}_n + E^{(1)}_{ 2 \rightarrow 3} = \frac{2- F_0
+ 2F^2_0}{3}.
\end{eqnarray}
The efficiency of our MEPP $E_o^{(1)}$ and the maximal value of that
from the identity-combination items for three-qubit systems
$E_n^{(1)}$ are shown in Fig.\ref{fig5_efficiency}(a). Also, we give
the efficiency that the three parties obtain three-electron
entangled systems from two-electron entangled subsystems
$E^{(1)}_{2\rightarrow  3}$ in Fig.\ref{fig5_efficiency}(a).

The fidelity of the two-electron subsystems obtained from
cross-combinations without two-electron entanglement purification is
\begin{eqnarray}
F^{(1)}_{2} &=& \frac{2F_0F_1}{2F_0F_1 + 2F_1^2 }=\frac{F_0}{F_0 +
F_1}=\frac{3F_0}{1 + 2F_0}.
\end{eqnarray}
The fidelity of the three-electron systems obtained from
two-electron subsystems with entanglement link is
\begin{eqnarray}
F^{(1)}_{2\rightarrow 3} &=& \frac{F^2_0}{(F_0+F_1)^2
}=\frac{9F^2_0}{1 + 4F_0 + 4F^2_0}.
\end{eqnarray}
The relation of the three fidelities $F^{(1)}_n$, $F^{(1)}_2$, and
$F^{(1)}_{2\rightarrow 3}$ is shown in Fig.\ref{fig5_efficiency}(b).

\begin{figure}[!h]
\begin{center}
\includegraphics[width=8cm,angle=0]{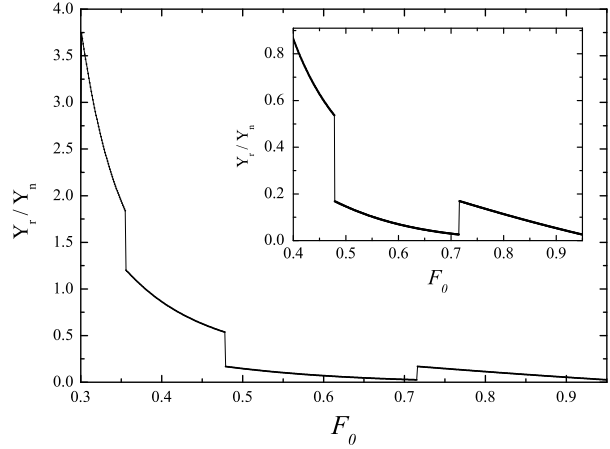}
\caption{The rate of the yield from our recycling thee-electron EPP
$Y_r$ to that from our normal three-electron EPP $Y_n$ with the
threshold value $F_{thr}=0.95$. In order to see the contribution of
our recycling EPP in a clear way, we use an insert to show the rate
for the original fidelity $F_0$ from $0.4$ to $0.95$.
}\label{fig6_yield}
\end{center}
\end{figure}

From Fig.\ref{fig5_efficiency}, one can see that our MEPP is more
efficient than the conventional MEPPs, especially in the case that
the original fidelity $F_0$ is not big. On the other hand, the
fidelity $F^{(1)}_{2\rightarrow 3}$ is smaller than $F^{(1)}_n$
although they both are larger than the original fidelity $F_0$ when
$F_0>\frac{1}{4}$.  $F^{(1)}_2$ is larger than $F^{(1)}_n$ when
$F_0<\frac{1}{2}$ and it is smaller than $F^{(1)}_n$ when
$F_0>\frac{1}{2}$. If three parties first run the two-electron EPP
$n$ times and then produce some three-electron systems with
entanglement link from high-fidelity two-electron subsystems, the
fidelity $F^{(n)}_{2\rightarrow 3} =
\frac{F_{0}^{2^{n+1}}}{(F_{0}^{2^{n}} + F_1^{2^n} )^2}$ can be
improved to be larger than $F^{(1)}_n$.

In order to show the contribution of the part from our recycling EPP
clearly, we calculate the rate of its yield $Y_r$ to that from our
normal EPP $Y_n$ for three-electron systems with the threshold value
$F_{thr}=0.95$, shown in Fig.\ref{fig6_yield}. From this figure, one
can see that the contribution of our recycling three-electron EPP is
larger than that of our normal EPP if the original fidelity $F_0$ is
smaller than $0.38$. When $F_0$ is no more than $0.478$, the
contribution of our recycling EPP is considerable. When $F_0$ is
larger than $0.716$, the number that the parties need to repeat
their EPP for obtaining three-electron systems with the fidelity
larger than the threshold value $F_{thr}=0.95$ from two-electron
subsystems is reduced to one, which increases the rate of the
contribution of our recycling EPP. As our three-electron EPP
contains two parts, that is, our normal EPP and our recycling EPP,
no matter what the original fidelity $F_0$ is, the yield of our
three-electron EPP is larger than that from conventional MEPPs as
the latter is just the part discarded in all conventional MEPPs
\cite{Murao,Horodecki,Yong,shengepjd,shengpla}.

\section{discussion and summary}

We have fully described the principle of our efficient
three-electron EPP for GHZ states. It is not difficult to prove that
our efficient EPP works for $N$-electron systems in a GHZ state. The
GHZ state of a multipartite entangled system composed of $N$
electrons can be described as
\begin{eqnarray}
|\Phi^{+}_0\rangle_N=\frac{1}{\sqrt{2}}(|\uparrow\uparrow\cdots
\uparrow\rangle + |\downarrow\downarrow\cdots
\downarrow\rangle)_{A,B,\cdots, Z}. \label{state2}
\end{eqnarray}
Here the subscripts $A$, $B$, $\cdots$, and  $Z$ represent the
electrons belonging to the parties Alice, Bob, $\cdots$, and Zach,
respectively. Certainly, there are another $2^N-1$ GHZ states for an
$N$-qubit system and can be written as
\begin{eqnarray}
|\Phi^{+}_{ij\cdots k}\rangle_N=\frac{1}{\sqrt{2}}(|ij\cdots
k\rangle + |\bar{i}\bar{j}\cdots \bar{k}\rangle)_{AB\cdots Z}
\end{eqnarray}
and
\begin{eqnarray}
|\Phi^{-}_{ij\cdots k}\rangle_N=\frac{1}{\sqrt{2}}(|ij\cdots
k\rangle - |\bar{i}\bar{j}\cdots \bar{k}\rangle)_{AB\cdots
Z}.\end{eqnarray} Here $\bar{i}=1-i$, $\bar{j}=1-j$,  $\bar{k}=1-k$,
and $i,j,k\in \{0,1\}$. $\vert 0\rangle\equiv \vert \uparrow\rangle$
and $\vert 1\rangle\equiv \vert \downarrow\rangle$. For correcting
the bit-flip errors in $N$-electron entangled quantum systems, we
can also divide the whole entanglement purification into two parts.
One is our normal $N$-electron entanglement purification and the
other is our recycling entanglement purification with entanglement
link from subsystems. Our normal entanglement purification for
$N$-electron entangled quantum systems with bit-flip errors is
similar to that for three-electron entangled quantum systems. We
should only increase the number of  the PCDs and the Hadamard
operations shown in Fig. \ref{fig2_MEEPP}. Let us assume that the
ensemble of $N$-electron systems after the transmission over a noisy
channel is in the state
\begin{eqnarray}
\rho_N &=& f^{'}_0|\Phi^+_0\rangle_N\langle \Phi^+_0| + \cdots +
f^{'}_{ij\cdots k} |\Phi^+_{ij\cdots k}\rangle_N\langle
\Phi^+_{ij\cdots k}| +   \cdots  +
f^{'}_{2^{N-1}-1}|\Phi^+_{2^{N-1}-1}\rangle_N\langle
\Phi^+_{2^{N-1}-1}|. \label{stateNm}
\end{eqnarray}
Here $f^{'}_{ij\cdots k}$ presents the probability that an
$N$-electron system is in the state $|\Phi^+_{ij\cdots k}\rangle_N$
and
\begin{eqnarray}
f^{'}_0 + \cdots + f^{'}_{ij\cdots k} + \cdots +
f^{'}_{2^{N-1}-1}=1.
\end{eqnarray}
In our normal $N$-electron entanglement purification for a pair of
systems, the parties will keep the identity-combination items
 $|\Phi^+_0\rangle_N \otimes |\Phi^+_0\rangle_N$, $\cdots$,
$|\Phi^+_{ij\cdots k}\rangle_N \otimes |\Phi^+_{ij\cdots
k}\rangle_N$, $\cdots$, and $|\Phi^+_{2^{N-1}-1}\rangle_N \otimes
|\Phi^+_{2^{N-1}-1}\rangle_N $ with the probabilities $f^{'2}_0$,
$\cdots$, $f_{ij\cdots k}^{'2}$, $\cdots$, and $f_{2^{N-1}-1}^{'2}$,
respectively. That is, they keep the instances in which they all
obtain the even parity and those in which they all obtain the odd
parity with their PCDs. With the similar process to the case for
three-electron systems, the parties can obtain a new  $N$-electron
system which is in the states $\vert
\Phi_0^+\rangle_{N}=\frac{1}{\sqrt{2}}(\vert HH\cdots H\rangle   +
\vert VV\cdots V\rangle)_{A_1,B_1,\cdots, Z_1}$, $\cdots$, $\vert
\Phi_{ij\cdots k}^+\rangle_{N}=\frac{1}{\sqrt{2}}(|ij\cdots k\rangle
+ |\bar{i}\bar{j}\cdots \bar{k})_{A_1,B_1,\cdots, Z_1}$, $\cdots$,
and $\vert \Phi_{2^{N-1}-1}^+\rangle_{N}= \frac{1}{\sqrt{2}}(\vert
HV\cdots V\rangle  + \vert VH\cdots H\rangle)_{A_1,B_1,\cdots, Z_1}$
with the probabilities $ f^{'2}_0$, $\cdots$, $ f_{ij\cdots
k}^{'2}$, $\cdots$, and $f_{2^{N-1}-1}^{'2}$, respectively. That is,
the parties can obtain a new ensemble of $N$-electron systems
$\rho'_N$ with the fidelity $f''_0=\frac{f^{'2}_0}{f_0^{'2} + \cdots
+ f_{ij\cdots k}^{'2} + \cdots + f_{2^{N-1}-1}^{'2}}$ from the
original ensemble in the state $\rho_N$. Our recycling EPP is used
to distill some $N'$-electron subsystems ($2 \leq N' < N$) from the
cross-combination items $|\Phi^+_{lr\cdots q}\rangle_N \otimes
|\Phi^+_{ij\cdots k}\rangle_N$ and $ |\Phi^+_{ij\cdots k}\rangle_N
\otimes |\Phi^+_{lr\cdots q}\rangle_N$ ($l,r,\cdots, q \in \{0,1\}$
and $l\neq i$, $r\neq j$, $\cdots$, or $q\neq k$). Its process is
also similar to the case with three-electron systems. The more the
number of the electrons in each system, the more the kinds of the
entanglement purification with entanglement link.

Compared with the conventional MEPPs
\cite{Murao,Horodecki,Yong,shengepjd,shengpla}, the present MEPP
contains two parts. One is our normal MEPP which is similar to the
conventional MEPPs as the high-fidelity $N$-electron systems are
directly obtained from the identity-combination items
$|\Phi^+_0\rangle_N \otimes |\Phi^+_0\rangle_N$, $\cdots$,
$|\Phi^+_{ij\cdots k}\rangle_N \otimes |\Phi^+_{ij\cdots
k}\rangle_N$, $\cdots$, and $|\Phi^+_{2^{N-1}-1}\rangle_N \otimes
|\Phi^+_{2^{N-1}-1}\rangle_N $. However, as the cross-combination
items $|\Phi^+_{lr\cdots q}\rangle_N \otimes |\Phi^+_{ij\cdots
k}\rangle_N$ and $ |\Phi^+_{ij\cdots k}\rangle_N \otimes
|\Phi^+_{lr\cdots q}\rangle_N$ ($l,r,\cdots, q \in \{0,1\}$ and
$l\neq i$, $r\neq j$, $\cdots$, or $q\neq k$) can not be used to
obtain high-fidelity $N$-electron systems directly, they are
discarded in the conventional MEPPs. In our recycling EPP, the
second part of our efficient MEPPs, the parties distill some
subsystems with a high fidelity from the cross-combination items.
With entanglement purification on the subsystems and the
entanglement production based on local entanglement link, the
parties can obtain some additional yield of $N$-electron systems
with the fidelity higher than the threshold value, which makes our
MEPP more efficient than the conventional MEPPs.

The PCD is the key element in our high-yield MEPP and charge
detection plays a crucial role in constructing the PCD for the spins
of two electrons. Charge detection has been realized by means of
point contacts in a two-dimensional electron gas. For instance,
Field \emph{et al.} \cite{cd} used the effect of the electric field
of the charge on the conductance of an adjacent point contact to
realize the charge detection in 1993. Elzerma \emph{et al.}
\cite{experiment1} reported their experimental results that the
current achievable time resolution for charge detection is $\mu s$
in 2004. Trauzettel \emph{et al.} \cite{parity} also proposed a
realization of a charge-parity meter which is based on two double
quantum dots alongside a quantum point contact in 2006. Their
realization of such a device can be seen as a specific example of
the general class of mesoscopic quadratic quantum measurement
detectors which is investigated by Mao \emph{et al.} \cite{parity2}.
Moreover, recent studies showed that the interaction between the
polarizations of photons and the electron spins of quantum dots in
optical cavities can also be used to construct the PCD, as shown in
Refs.\cite{qndother1,qndother2,qndother3,qndother4,qndother5,qndother6}

In summary, we have proposed a high-efficiency MEPP for $N$-electron
systems in a GHZ state, resorting to the PCD based on charge
detection. It contains two parts. One is our normal MEPP with which
the parties can obtain a high-fidelity $N$-electron ensemble
directly. This part comes from the identity-combination items of a
pair of $N$-electron systems,  similar to conventional MEPPs but
with a higher efficiency. The cross-combination items, which are
discarded in all existing conventional MEPPs, can be used to distill
some $N'$-electron subsystems ($2 \leq N' < N$) by measuring the
electrons with potential errors in our recycling MEPP, the second
part of our high-yield MEPP. Our normal MEPP has a higher efficiency
than the MEPP for a Werner-type state with perfect CNOT gates
\cite{Murao}. Especially, it doubles the efficiency of the MEPP with
QNDs based on cross-Kerr nonlinearity in Ref. \cite{shengepjd} and
the MEPP for electronic systems \cite{shengpla}. In our recycling
MEPP, the parties in quantum communication can produce some
high-fidelity $N$-electron systems with entanglement link based on
the parity-check detection. Combining the second part of our MEPP
with the first one, the present MEPP has a higher efficiency and
yield than the conventional MEPPs largely.

\section*{ACKNOWLEDGEMENTS}

This work is supported by the National Natural Science Foundation of
China under Grant Nos. 10974020 and 11174039, NCET-11-0031, and the
Fundamental Research Funds for the Central Universities.

\end{document}